\documentclass[twocolumn,showpacs]{revtex4}
\usepackage{amssymb,amsmath,amsfonts}
\usepackage{graphicx,graphics}
\usepackage{color}

\usepackage{framed}
\newcommand{\rem}[1]{}
\DeclareMathAlphabet{\mathbi}{OML}{cmm}{b}{it} 
\newcommand{\non}{\nonumber}

\newcommand{\bx}{\mathbi{x}}

\newcommand{\bel}{\begin{equation}\label}
\newcommand{\ee}{\end{equation}}
\newcommand{\beq}{\begin{eqnarray}\label} 
\newcommand{\eeq}{\end{eqnarray}} 
\newcommand{\bc}{\begin{center}} 
\newcommand{\ec}{\end{center}} 
\newcommand{\ben}{\begin{enumerate}}
\newcommand{\een}{\end{enumerate}}
\newcommand{\bit}{\begin{itemize}}
\newcommand{\eit}{\end{itemize}}

\newcommand{\bD}{\mbox{\boldmath$\mathcal{D}$}}

\newcommand{\bdB}{\mathbi{\mathcal{B}}}

\newcommand{\bu}{\mbox{\boldmath$u$}}

\newcommand{\bJ}{\mathbi{\mathcal{J}}}
\newcommand{\bU}{\mathbi{\mathcal{U}}}

\newcommand{\bom}{\mbox{\boldmath$\omega$}}

\begin{document}
%%%%%
\title{Quasi-conservation laws for compressible 3D Navier-Stokes flow}
%\par\medskip
\author{J. D. Gibbon$^{1,2}$ and D. D. Holm$^1$}
%\par\vspace{1mm}
\affiliation{
$^1$Department of Mathematics, Imperial College London SW7 2AZ, UK\\
$^2$Isaac Newton Institute for Mathematical Sciences, Cambridge CB3 0EW, UK}
\email{j.d.gibbon@ic.ac.uk; d.holm@ic.ac.uk}
\homepage{http://www2.imperial.ac.uk/~jdg; http://www2.imperial.ac.uk/~dholm/}

\begin{abstract}
We formulate the quasi-Lagrangian fluid transport dynamics of mass density $\rho$ and the projection $q=\bom\cdot\nabla\rho$ of the vorticity $\bom$ onto the density gradient, as determined by the 3D compressible Navier-Stokes equations for an ideal gas, although the results apply for an arbitrary equation of state. It turns out that the quasi-Lagrangian transport of $q$ cannot cross a level set of $\rho$. That is, in this  formulation, level sets of $\rho$ (isopychnals) are impermeable to the transport of the projection $q$.
\end{abstract}
\maketitle
\vspace{-3mm}
%%%%%%%%%%%%%%%%%%%

The aim of this note is to formulate the fluid conservation  dynamics for mass density $\rho$ and the projection $q=\bom\cdot\nabla\rho$ of the vorticity $\bom={\rm curl}\,\bu$ onto the mass density gradient, as determined by the 3D compressible Navier-Stokes equations  \cite{LLhydro}
\beq{comp1}
\rho\,\frac{D\bu}{Dt}  =  
\mu\Delta\bu\ -\nabla \varpi 
\,,&& \varpi = p - (\mu/3+\mu^v){\rm div}\,\bu
\,,\quad
\\
\frac{D\rho}{Dt} + \rho\,\mbox{div}\,\,\bu = 0
\,,&& 
\frac{D~}{Dt} = \partial_{t} + \bu\cdot\nabla
\,,\label{comp2}
\\
c_v\frac{D\theta}{Dt} = \frac{p}{\rho}\,\mbox{div}\,\,\bu + Q 
\,,&& 
p = R \rho \theta 
\,.
\label{comp3}
\eeq
Here, $\bu$ denotes the spatial fluid velocity, $\mu$ is the shear viscosity and $\mu^v$ is the volume viscosity, both of which are  taken as constitutive constants of the fluid. 

For definiteness, we have chosen an ideal gas equation of state to relate pressure, $p$, temperature, $\theta$, and mass density, $\rho$. In addition, $R$ is the gas constant,  $c_v$ the specific heat is constant and $Q$ is the heating rate, which we may assume is known. 
The pressure depends on the two thermodynamic variables $\rho$ and $\theta$. It is noteworthy that the dynamics of the temperature and the choice of equation of state in (\ref{comp3}) will not affect our considerations below of the transport dynamics of the projection $q=\bom\cdot\nabla\rho$. 
This means the geometric considerations that follow are  universal for any viscous compressible fluid flow. 

According to equations (\ref{comp1}) and (\ref{comp2}) 
the vorticity $\bom = \mbox{curl}\,\bu$ evolves according to a stretching and folding equation
\begin{align}\label{Dom}
\partial_{t}\bom &- \mbox{curl}\,(\bu\times\bom) 
\\
& =
\mu\rho^{-1}\Delta\bom
+
\nabla\rho^{-1}\times\left[\mu\Delta\bu 
-\nabla \left(\varpi - \frac{u^2}{2}\right)\right]
,\non
\end{align}
driven by the right hand side.
The form of the last term invites a projection against $\nabla\rho$.
By the product rule, 
the projection $q=\bom\cdot\nabla\rho$ satisfies a geometric relation reminiscent of
Ertel's Theorem \cite{Ertel42}
\begin{align}
\frac{Dq}{Dt} = \left(
\frac{D\bom}{Dt} 
- \bom\cdot\nabla\bu\right)\cdot \nabla \rho 
+ \bom\cdot\nabla\left(\frac{D\rho}{Dt}
\right)
\label{q-calc1}
\end{align}
Ertel's Theorem, important in atmospheric dynamics, has the same form as (\ref{q-calc1}), with mass density replaced by potential temperature. Substituting mass conservation (\ref{comp2}) and Navier-Stokes vorticity dynamics (\ref{Dom}) into the purely geometric equation (\ref{q-calc1}) yields the following equation for $q$ written in a suggestive divergence form,
\begin{align}\,\hspace{-1mm}
\partial_t q + \mbox{div}\,q\bu
+ \mbox{div}\big(\bom\rho\,\mbox{div}\,\bu
- \mu\Delta\bu\times\nabla(\ln\rho)\big) = 0\,.
\label{q-calc2}
\end{align}

Now we apply an observation of Haynes and McIntyre \cite{HMc87} that 
first arose in atmospheric physics and allows one to define a current density $\mathbf{J}$. Assuming that solutions exist for equations (\ref{comp1})--(\ref{comp3}), as
\bel{bigU}
\mathbf{J} = q\bu 
+ \bom\rho\,\mbox{div}\,\bu 
- \mu\Delta\bu\times\nabla(\ln\rho)
+ \nabla\phi\times \nabla f(\rho)\,,
\ee
where $\phi$ is an undetermined gauge potential and $f$ is an arbitrary differentiable function of $\rho$. 
Then (\ref{q-calc2}) and (\ref{comp2}) can be rewritten in the \emph{quasi-conservative} form
\bel{q1d}
\partial_{t}q + \mbox{div}\,\mathbf{J} = 0
\quad\hbox{and}\quad
q\,\partial_{t}\rho + \mathbf{J}\cdot\nabla\rho = 0\,.
\ee
The relation $ \mathbf{J}\cdot\nabla\rho=q\,\mbox{div}\,\rho\bu $ allows zero projection, $q=0$, by the second equation in (\ref{q1d}). Thus, the projection $q$ may vanish anywhere in the flow, but it cannot be maintained, because $\mbox{div}\,\mathbf{J} \ne0$. Together, the equations in (\ref{q1d}) imply a family of conserved quantities, since
\bel{cons}
\partial_{t} (q\Phi'(\rho)) +  \mbox{div}\,(\mathbf{J}\Phi'(\rho))=0
\,,\ee
for any function $\Phi'(\rho)=d\Phi/d\rho$.

The  conserved  densities $q\Phi'(\rho)={\rm div}(\Phi(\rho)\bom)$ in (\ref{cons}) possess quite  different flow properties from those  of  mass, energy  and momentum. The question of the physical interpretation  of  these quantities  may now be  raised. We know that  
equations (\ref{q1d}) are purely kinematic, because the projection taken against $\nabla\rho$ removed any dependence on the dynamics of the temperature and the choice of equation of state in (\ref{comp3}). 
Nonetheless, equations (\ref{q1d})  have been derived \emph{without approximation} from the Navier-Stokes  fluid equations for compressible motion and mass transport. 

Moreover, their analogues also occur  
and have been found useful in  other areas  of  fluid dynamics, particularly in the dynamics of the ocean and atmosphere  \cite{HMR85,HMc87}. (In the oceanic context, $\rho$ denotes buoyancy and the motion is usually taken to be incompressible.)

Indeed, a similar calculation may be performed to derive the quasi-conservative form in (\ref{q1d}) of the dynamics of the projection of vorticity on temperature gradient, $q'=\bom\cdot\nabla\theta$ from equation (\ref{q-calc1}) with $\nabla\rho$ replaced by $\nabla\theta$. The projection $q'$ with $\theta$ denoting the potential temperature is known as the \emph{potential vorticity density} and has been useful in guiding thinking about nonlinear convective processes in atmospheric science, \cite{HMR85,HMc87}. 

In addition, the corresponding forms of (\ref{bigU})--(\ref{q1d}) may be written in the absence of viscosity. One hopes that with the proper interpretation equations (\ref{q1d}) may also be useful in considerations of compressible Navier-Stokes flows. 

Following \cite{HMR85,HMc87} in the atmospheric context for which the projection $q$ is replaced by potential vorticity $q'$, we propose to interpret equations (\ref{bigU})--(\ref{q1d}) locally as a type of \emph{impermeability theorem} for the \emph{quasi-Lagrangian} transport of the projection $q$ and the mass density $\rho$ by a pseudo-velocity $\bU=\mathbf{J}/q$. 
Here, impermeability means that 
the projection $q= \bom\cdot\nabla\rho$ cannot be transported by pseudo-velocity $\bU$ 
\footnote{By the definition of $\mathbf{J}$ in (\ref{bigU}) the pseudo-velocity $\bU=\mathbf{J}/q$ is determined only up to the addition of the curl of a vector proportional to $\nabla\rho$.}
across level sets of the mass density, $\rho$.  
This is remarkable because the two quantities $q$ and $\rho$  are transported by \emph{different velocities}. 

The quasi-conservative form of compressible Navier-Stokes fluid dynamics in (\ref{q1d}) has apparently not appeared 
previously in the literature, and we hope that this formulation may find some future use, perhaps in analogy with the use of potential vorticity in the atmospheric context. 

In particular, these equations may be useful in the study of stretching and folding in compressible fluid flows, just as has been recently investigated in the atmospheric context in \cite{JDGDD10,JDGDD11}.  Although previous studies of compressible flows have often focused on shock formation, we shall close this note with a remark about how the gradient $\nabla q$ of the projection $q=\bom\cdot\nabla\rho$ participates in stretching, folding and expansion of higher-order gradients  in compressible Navier-Stokes fluid flows. In preparation, we rewrite Equations (\ref{comp2}) and (\ref{q1d}) in terms of the pseudo-velocity $\bU=\mathbf{J}/q$ as
\bel{q1d2}
\partial_{t}q + \mbox{div}\,\left(q\,\bU\right) = 0
\qquad
\partial_{t}\rho + \bU\cdot\nabla\rho = 0
\,.\ee
Note however that  $\mbox{div}\,\bU \ne 0$. We define 
\bel{Bdef}
\bdB = \nabla q\times\nabla \rho
\,,\ee
and discover from a direct computation that
$\bdB$ satisfies
\bel{ceeqn1}
\partial_{t}\bdB - \mbox{curl}\,(\bU\times\bdB) = \bD
\,,\ee
where $\bD = - \nabla(q\,\mbox{div}\,\bU)\times\nabla\rho$. The proof of the corresponding relation in an atmospheric physics context can be found in \cite{JDGDD10,JDGDD11}. 

Based on the transport velocity defined in (\ref{bigU}), the left hand side of (\ref{ceeqn1}) makes it clear that the 
vector $\bdB$ undergoes the same type of stretching, folding and expansion processes driven by the $\bD$-vector on the right hand side as occurs in the vorticity equation (\ref{Dom}). 
Explicitly, the stretching, folding and expansion processes are given by
\begin{align}
\partial_{t}\bdB + 
\underbrace{\
\bU\cdot \nabla \bdB - \bdB\cdot \nabla \bU
+ \bdB {\,\rm div\,}\bU 
\
}_{\hbox{\it stretch, fold \& expand $\bdB$}}
 =\ \bD\,.\vspace{2mm}
 \label{BeqnSF}
\end{align}
This is remarkable, because $\bdB = \nabla q\times\nabla\rho$ contains information not only about $\nabla\bom$ but also about $\nabla\rho$ and even $\nabla\nabla\rho$. Thus, the stretching and folding in the original vorticity equation (\ref{Dom}) compounds itself in the same form in the $\bdB$-equation (\ref{ceeqn1}), but with higher spatial derivatives.

\par\smallskip\noindent
{\bf Acknowledgements.} We are grateful to B. J. Hoskins and the anonymous referees for suggestions that have improved and simplified this article. The work by DDH was partially supported by an Advanced Grant from the European Research Council.

%%%%%%%%%%%%%%%%%%%%%%%%%%%

\end{document}